\documentclass[12pt]{article}
\usepackage{epsfig,enumerate,verbatim}
\usepackage{graphicx}
\begin{document}

\begin{center}
{\large \bf Universality in an Information-theoretic Motivated Nonlinear Schrodinger Equation}
\end{center}
\vspace{0.1in}

\begin{center}

{R. Parwani\footnote{Email: parwani@nus.edu.sg} and G. Tabia}

\vspace{0.3in}

{Department of Physics,\\}
{National University of Singapore,\\}
{Kent Ridge,\\}
{ Singapore.}

\vspace{0.3in}
31 July 2006; Revised Feb 05 2007.
%\maketitle
\end{center}
\vspace{0.1in}
\begin{abstract}

Using perturbative methods, we analyse a nonlinear generalisation of Schrodinger's equation that had previously been obtained through information-theoretic arguments. We obtain analytical expressions for the leading correction, in terms of the nonlinearity scale, to the energy eigenvalues of the linear Schrodinger equation in the presence of an external potential and observe some generic features. In one space dimension these are: (i) For nodeless ground states, the  energy shifts are subleading in the nonlinearity parameter compared to the shifts for the excited states, (ii) the shifts for the excited states are due predominantly to contribution from the nodes of the unperturbed wavefunctions and (iii) the energy shifts for excited states are positive for small values of a regulating parameter and negative at large values, vanishing at a universal critical value that is not manifest in the equation. Some of these features hold true for higher dimensional problems. We also study two exactly solved nonlinear Schrodinger equations so as to contrast our observations. Finally, we comment on the possible significance of our results if the nonlinearity is physically realised. 

\end{abstract}
%\pacs{}

\vspace{0.5in}

\section{Introduction}

Various nonlinear extensions of Schrodinger's equation have been proposed 
\cite{probe} over the years as possible generalisations of the linear
evolution of the original theory. Although several low-energy experiments have placed
very small upper bounds \cite{bdds} on the proposed extensions, there is
still the possibility that quantum mechanics might have to be modified at
high energies or short distances \cite{RP1} where the structure of spacetime is expected to be different \cite{Garay}. 

However in this paper we remain within the non-relativistic realm so as to
explore in more detail the properties of one particular nonlinear extension that was motivated in \cite{RP2} by
maximum uncertainty arguments \cite{fried1,P1} similar to those used in
statistical mechanics \cite{Jay}. In higher than one space dimension the
equation of Ref.\cite{RP2} was not rotationally invariant, motivating a suggestive link between spacetime symmetries and
quantum linearity. Some implications of such a connection for phenomenology were
discussed heuristically in \cite{RP2,RP1}.

In Ref.\cite{tan} some exact nonperturbative solutions of the abovementioned 
equation were obtained, displaying  intriguing and novel features that are probably related to the unusual 
structure of that equation. Indeed, it was hinted in \cite{tan} that the equation might also be interesting as an effective equation in other domains of physics, such as nonlinear optics \cite{sulem},
rather than its original intention in Ref.\cite{tan}.

Here we investigate how the nonlinearity of that equation perturbs the energy 
spectrum of the usual linear Schrodinger
equation. Since simple estimates already indicate that the size of the
nonlinearity scale must be tiny for it to be consistent with phenomenology 
\cite{RP2}, we shall use standard first order perturbation theory for our study. Our primary aim
here is not to confront empirical data but to uncover further  properties
of the nonlinear equation. As we shall see, even at the perturbative level
the equation of Ref.\cite{RP2} has a rather surprising character. In particular, we find a  
universal critical point of the theory that is not at all obvious from the equations of motion. 

In the next section we outline our perturbative scheme and then illustrate it with some 
numerical results in Section (3).  A general analytical investigation is next conducted in Section(4) 
to extract and explain the features mentioned in the abstract. 
In Section (5) we contrast the perturbative properties of the
nonlinear equation with those of two other exactly solved nonlinear
Schrodinger equations. The concluding section summarises the main lessons
and discusses some implications. The appendices contain additional derivations.

\section{Perturbative framework}

Let us focus first on the nonlinear equation for a single particle in one
space dimension that was derived in Ref.\cite{RP2}, 
\begin{equation}
i \hbar {\frac{\partial \Psi }{\partial t}} = - {\frac{{\hbar}^2 }{2m}} {%
\frac{\partial^2 \Psi }{\partial x^2}} + V(x) \Psi + F(p) \Psi \, ,
\label{nsch1}
\end{equation}
with $p(x,t) = \Psi^{\star}(x,t) \Psi(x,t)$ the conserved probability
density and 
\begin{eqnarray}
F(p) &\equiv& Q_{nl} - Q \, ,  \label{FF}
\end{eqnarray}
where 
\begin{equation}
Q_{nl} = {\frac{{\hbar}^2 }{4 m L^2 \eta^4 }} \left[ \ln {\frac{p }{%
(1-\eta) p + \eta p_{+} }} + 1 - {\frac{(1-\eta) p }{(1-\eta) p + \eta p_{+}}%
} - {\frac{\eta p_{-} }{(1-\eta) p_{-} + \eta p}} \right] \,  \label{Q2}
\end{equation}
is a regularised nonlinear ``quantum potential". The parameter $\eta$ takes
values $0 < \eta < 1$, its crucial role being to regulate potential
singularities where $p(x)$ vanishes. We have used the notation 
\begin{eqnarray}
p_{\pm}(x) &= p(x \pm \eta L) \, . \label{pm}
\end{eqnarray}
Note that if $\Psi$ is any solution of the equation, then so is $\lambda
\Psi $ for an arbitrary constant $\lambda$, so we may re-normalise states
freely. The nonlinearity is characterised by the length scale $L$, in terms
of which one may perform a formal expansion of (\ref{Q2}), 
\begin{equation}
Q_{nl} \to Q \equiv - {\frac{{\hbar}^2 }{2m}} {\frac{1 }{\sqrt{p}}} {\frac{%
\partial^2 \sqrt{p} }{\partial x^2}} \, , \label{pot1}
\end{equation}
with a remainder of $O(L)$.

Let $\Psi = e^{-iEt / \hbar} \phi(x)$ be the energy eigenstates of the usual
linear Schrodinger equation for a given external potential $V(x)$. Assuming
that the spectrum deforms continuously as the nonlinearity $F$ is turned on,
then to leading order the corrected energies are given by first-order
perturbation theory, 
\begin{eqnarray}
E_{exact} &=& E + \delta E \, , \\
\delta E &=& \int_{-\infty}^{\infty} dx \ \phi^{\star} F(\phi) \phi \,\, .
\label{delta}
\end{eqnarray}
Note that $F$ is evaluated using the unperturbed wavefunctions and so from
now on $p$ will refer to $\phi^{\star}(x) \phi(x)$.

We may trust first-order perturbation theory when the nonlinearity is small.
The relevant dimensionless expansion parameter is $L/a$ where $a$ is a typical scale
in the linear theory, such as the deBroglie wavelength. As equations (\ref{Q2},\ref{pm})
indicate, the expression (\ref{delta}) is actually a complicated function of $L/a$
from which the leading behaviour must be extracted. We will discard
subleading terms from (\ref{delta}) as they will be of the same order as
second-order perturbation theory corrections, which we do not study here.

For the problems we will study, the unperturbed wavefunctions $\phi (x)$ are
parity eigenstates so that $p(x)=p(-x)$. Changing variables $x\rightarrow -x$
in (\ref{delta}) and using the parity invariance of $p$ shows that 
\begin{equation}
\delta E(L)=\delta E(-L)\,. \label{parity}
\end{equation}
That is, when we allow the parameter $L$ to take negative values, then although equation (\ref{nsch1}) is not invariant under 
 $L\rightarrow -L$, yet the first-order energy shifts are. Therefore if $%
\delta E(L)$ were an analytic function of $L$, one would have concluded that 
\begin{equation}
\delta E(L)\sim O(L^{2})+O(L^{4})+...\,
\end{equation}
as the $O(L^{0})$ term vanishes by construction, see (\ref{pot1}). In
reality however, $\delta E(L)$ is generically non-analytic! To see this,
consider the naive series expansion of the integrand in (\ref{delta}). It
results in the formal expression \newline
\begin{eqnarray}
\delta E(L) &\propto& L^2 \eta ^{2}\int_{\infty
}^{\infty }{\frac{dx}{p^{3}}}\;  \mbox{[} 6(2-3\eta )^{2}(p^{\prime })^{4}\ -\
12(3-8\eta \ +\ 6\eta ^{2})p(p^{\prime })^{2}p^{\prime \prime }  \nonumber \\
&&\hspace{2cm}\;\;\;+\ 4p^{2}p^{\prime }p^{\prime \prime \prime }\ +\
p^{2}(3(p^{\prime \prime })^{2}-2pp^{\prime \prime \prime \prime }) \mbox{]} \,
\label{formal}
\end{eqnarray}
which is ill-defined because of the singularities that occur where $p(x)$
vanishes, that is where the unperturbed wavefunction has nodes. Thus one may
conclude $\delta E(L)\sim O(L^{2})$ only for nodeless states, which are
typically only the ground states of a system.

Since excited states of the unperturbed theory have nodes, we cannot use (%
\ref{formal}) for them. In the next section we perform a numerical investigation
of expression (\ref{delta}) and then return in Section (4) to a more general
analytical investigation that explains the various observed results, such as $\delta E(L)
\sim O(|L|)$ for states with nodes.

\section{Numerical Investigation}
The purpose of the numerical study is two-fold. Firstly it helps us uncover some interesting features of the complicated nonlinearity (\ref{FF}) and so guides us in the later, more general, analytical investigations. Secondly, it will provide us with important checks on the analytical derivations of Section (4) and in particular answer the question of how small, numerically, the perturbative parameter $L/a$ has to be in the analytical expressions.

As convergence near the end-points $\eta=0,1$ is slow, we integrate (\ref
{delta}) numerically at the symmetric point $\eta =1/2$, defering a
discussion of other $\eta$ values to Sect.(4). Although phenomenologically
one expects $L/a$ to be tiny \cite{RP2}, we study much larger values $%
\sim 10^{-3}$ for computational efficiency. However we do demand $\delta E /
E \sim 10^{-2}$ or smaller so as to be safely in the perturbative regime.
For each $V(x)$, we obtain the leading dependence of $\delta E$ on $L/a$ and
the principal quantum number. The numerical results are then parametrised
using a best fit to simple analytical power law expressions.

The numerical work was performed with Mathematica \cite{Math} and the quoted numbers are
accurate to about the last digit.

In the numerical work we have set $L=1$ to define the reference units. Thus 
$1/a$ factors quoted below actually correspond to the dimensionless
quantity $L/a$. We have checked that the numerical results are invariant
under $L\rightarrow -L$ as required by the parity invariance argument of
Sect.(2).

\subsection{Infinite Well}

The infinite well with walls at $x=0$ and $x=a$ gives 
\begin{equation}
\phi_n (x) = \sqrt{{\frac{2 }{a}}} \sin {\frac{n \pi x }{a}}  \label{well}
\end{equation}
and unperturbed energies 
\begin{equation}
E_{n}^{0} = {\frac{\hbar^2 \pi^2 n^2 }{2 m a^2}} \, .
\end{equation}
In the presence of the nonlinearity the energies shift and are given to
leading order by 
\begin{equation}
E_n = E_{n}^{0} + \delta E_n \, .  \label{eng}
\end{equation}
It is convenient to define dimensonless quantities by dividing the above
equation by $\hbar^2 \pi^2 / 2m a^2$, 
\begin{equation}
\tilde{E}_n = n^2 + \delta \tilde{E_n} \, .  \label{eng-dim}
\end{equation}

For various fixed values of $n$, $1 \le n \le 50$, the energy shifts were
evaluated numerically for $1000 < a < 10,000$. Fig.(1) shows a log-log plot
for the $n=1$ case from which one deduces $\delta \tilde{E} = -0.99/a$. The
other $n$ values give similar plots, all indicating $\delta \tilde{E}
\propto -1/a$.

On the other hand, for various fixed $a$, an evaluation over the range $%
5\leq n\leq 50$ shows $\delta \tilde{E}\propto -n^{3}$. Re-inserting `$L$' we find, averaging the best fit for various $a$ values, 
\begin{equation}
\delta \tilde{E}_{n}=-1.03{\frac{n^{3}|L|}{a}}+O(L/a)^{2} \, . \label{well2}
\end{equation}
In Appendix A we will
explain this result analytically.

Notice that the correction (\ref{well2}) grows with $n$ and so at some large value of $n$ it is no longer small compared to the unperturbed value. This simply means that one must then go beyond first order perturbation theory. We discuss the possibilities in the concluding section.

\subsection{Simple Harmonic Oscillator (SHO)}

The potential is now $V(x)=kx^{2}/2$ giving the usual unperturbed
wavefunctions \cite{qmbook}
\begin{equation}
\phi _{n}(x)=\frac{1}{\sqrt{n!2^{n}}}\left( \pi a^{2}\right) ^{-1/4}\ H_{n} \left(%
\frac{x}{a} \right)\ \exp \left( -\frac{x^{2}}{2a^{2}}\right)
\end{equation}
and unperturbed energies 
\begin{equation}
E_{n}^{0}=(n+{\frac{1}{2}})\ \hbar \ \sqrt{{\frac{k}{m}}} \, .
\end{equation}
The linear length scale `$a$' in this problem is the deBroglie length $\hbar
^{1/2}/(km)^{1/4}$. We investigated the dimensionless energies shifts 
\begin{equation}
\delta \tilde{E}_{n}\equiv {\frac{\delta {E}}{\hbar }}\sqrt{\frac{m}{k}}
\label{eng-dim}
\end{equation}
numerically over the range, $0\leq n\leq 18$ and $100<a<1000$.

For the ground state, $n=0$, we find $\delta \tilde{E}\propto
-1/a^{2}$, which is a faster drop than seen for the infinite well. However
excited states have a similar behaviour in `$a$' to those of the infinite
well, with $\delta \tilde{E}\propto -1/a$ for any fixed $n$. The $n-$dependence for fixed $a$ is more complicated as indicated in
Fig.(2). In summary we find for the excited states, $n \ge 1$, 
\begin{equation}
\delta \tilde{E}_{n}=-0.26{\frac{n^{1.41}|L|}{a}}+O(L/a)^{2}\,.  \label{sho}
\end{equation}

It must be emphasized that the result (\ref{sho}) is a best fit to an assumed power law over the limited range investigated. However, independent analytical estimates in Appendix B do give a similar result over the same range.

\subsection{Hydrogen Atom}

We use the standard unperturbed wavefunctions as given, for example, in Ref.\cite{qmbook},
\begin{equation}
\psi _{nlm}(r,\theta ,\phi )=\sqrt{\left( \frac{2}{na}\right) ^{3}\frac{%
(n-l-1)!}{2n[(n+l)!]}}\rho ^{l}e^{-\rho /2}L_{n-l-1}^{2l+1}(\rho
)Y_{m}^{l}(\theta ,\phi ), \label{hyd}
\end{equation}
with $\rho =2r/na$, and the corresponding unperturbed energies  
\begin{equation}
E_{n}^{0}=-\frac{\hbar ^{2}}{2Ma^{2}n^{2}},
\end{equation}
where $M$ is the electron mass.

The three dimensional version of eq.(\ref{FF}) is  \cite{RP2}
\begin{eqnarray}
F(p) & \equiv & Q_{3} - Q \, , \label{3d} \\
Q_{3} &=& \sum_{i=1}^{3} {\frac{{\hbar}^2 }{4 M L^2 \eta^4 }}  \left[ \ln {p \over (1-\eta) p + \eta p_{+i} } + 1 - {(1-\eta) p \over (1-\eta) p + \eta p_{+i}} - {\eta p_{-i} \over (1-\eta) p_{-i} + \eta p} \right] \, ,  \nonumber \\
Q &=& -{{\hbar}^2 \over 8M}  \left[ {2 \partial_i \partial_i p \over p} - {\partial_i p \partial_i p \over p^2} \right] \, , \label{3dd}
\end{eqnarray}
with $i=1,2,3$ and $p_{\pm 1}(x)=p(x_1 \pm \eta L, x_2, x_3)$ and so on.
The Bohr radius defines $a=\hbar ^{2}/Me^{2}$ for this problem and the
dimensionless energy shifts are  
\begin{equation}
\delta \tilde{E}_{n}\equiv {\frac{\delta {E}}{\hbar ^{2}}(2M}a^{2}) \, .
\label{hatom-shift}
\end{equation}

Note that the nonlinearity breaks rotational invariance in the above expression (\ref{3d}) which is defined in the 
preferred Cartesian basis as discussed 
in Ref.\cite{RP2}. Thus the wavefunctions (\ref{hyd}) are first converted to the Cartesian basis for use in (\ref{3d}) but the final numerical integration was performed after converting back to
spherical coordinates. We used the built-in Monte Carlo subroutine in
Mathematica \cite{Math} for this case and investigated only a very limited range of
parameter values due to the time-intensive nature of the three dimensional
problem.

Although the pure Coulombic hydrogen atom states have a degenerate spectrum,
we still used the simple non-degenerate first order perturbation theory
formula for all states as our primary objective is to observe the effects of the 
nonlinearity on energy shifts. (Note also that  the energy shifts due to the
nonlinearity are expected to be much less than other effects, such as
relativity, that in reality lift the degeneracy of the unperturbed states.)

Consider first the zero angular momentum, $l=0$ states. For the ground
state, $n=1$ we found $\delta \tilde{E}\propto -1/a^{2}$ while for the $n=2,3$
excited states we have $\delta \tilde{E}\propto -1/a$. This dependence on $a$
is similar to that of the SHO. The dependence of
the energy shifts on the principal quantum number however appears to be much more
complicated than the earlier one-dimensional problems. Figure (3) plots $%
|\delta E/E_{0}|$ for $n\geq 2$.

For higher angular momentum states, there is a clear difference between the $%
n=l-1$ cases and $n\neq l-1$. For the former case we find $\delta 
\tilde{E}\propto -1/a^{2}$, a behaviour typical of nodeless states, while
for latter case we find the expected $\delta \tilde{E}\propto -1/a$ trend.
We explain the distinction between the two
cases in Sect.(4).

As for the dependence on the magnetic quantum number $m$ we do have the
expected invariance under $m\rightarrow -m$, but also find a mild dependence
of the energy shift for for different $m$ corresponding to the same $n,l$. For example,
on a log-log plot of $\left| \delta \tilde{E}\right| $ vs. $a$
for the $n=3,\ l=2,m=1,2$ states of the Hydrogen atom, the $m=2$ line has slope $%
-1.979(7)$ and intercept $-10.14(3)$ while the $m=1$ line has slope $%
-1.986(4)$ and intercept $-9.61(2)$.

\section{General Analytical Investigation}

Unless otherwise stated, in this section we discuss the nonlinear equation in the presence of a general smooth external potential $V(x)$ which for convenience we choose to be parity even, $V(x)=V(-x)$, so that the unperturbed states are parity eigenstates.
Since most studied potentials are parity even, that restriction is not unreasonable. However we emphasize that the key features of our result, such as eq.(\ref{EPX1}) below, follow from the structure of the nonlinearity (\ref{FF}): For example, without using parity eigenstates below, one still obtains similar results if instead of (\ref{FF}) one uses the $L \to -L$ symmetrised version \cite{RP2} of the nonlinearity.

\subsection{Nodeless States}

If $p(x)$ does not vanish in the region of integration, such as the ground
state of the SHO, one may use (\ref{formal}) to conclude that $\delta E \sim
O(L^2)$. Explicitly, we have for the $n=0$ SHO state, 
\begin{equation}
\delta \tilde{E} = {\frac{ \eta^2 ( 1 - \eta ) ( 1 - 3 \eta) }{4 }}
\left(\frac{ L }{a }\right)^2 + O(L^4)   \label{sho-o}
\end{equation}
which for $\eta =1/2$ is in excellent agreement with the leading result
extracted numerically in Section (3.2), $\delta \tilde{E} = -0.0156 L^2/a^2$. 
Equation (\ref{sho-o}) indicates a number of
interesting features: It vanishes both as $\eta \to 0$, which is the formal
linear limit of (\ref{nsch1}) and also as $L/a \to 0$ which is the physical
linear limit. $\delta \tilde{E}$ also vanishes at $\eta =1/3$ and $\eta =1$
but it is apparent from (\ref{formal}) that unlike the $\eta \to 0$ case the
other two critical values are dependent on $V(x)$.

We also note that $\delta \tilde{E}$ in (\ref{sho-o}) is positive for $\eta
< 1/3$ and negative for larger values. Such crossing behaviour will also be
seen below for excited states but, more remarkably, at a universal (that is, $V(x)$ independent) value of 
$\eta$.

The conclusion $\delta E\sim O(L)^{2}$ that we have drawn for nodeless
states from (\ref{formal}) is for smooth one-dimensional potentials. For
higher dimensions the conclusion is still true because of the form of (\ref
{3d}) but now one may encounter some nodes that are integrable, as in the
hydrogen atom case to be discussed in Sect.(4.3) below.

\subsection{Excited States in One Dimension}

When the unperturbed wavefunction $\phi(x)$ vanishes at a number of nodes
the formal $L$-expansion of the quantum potential $Q_{nl}$ as in (\ref{formal}) breaks down and so one has to proceed differently. Now, from the definition of $\delta E$  in eq.(\ref{delta}) we have
\begin{equation}
\delta E = \int p \ Q_{nl} - \int p \ Q \,. \label{diff}
\end{equation}
The second integral in (\ref{diff}) is independent of $L$ and it will cancel the $L$-independent piece of the first integral. So let us focus on the first integral in (\ref{diff}): Suppose first that $p(x)$ has exactly one node
at $x=x_1$. Since there are two widely separated length scales, $|L| \ll a $,
we may divide the integration region in the first term  of (\ref{diff}) into three parts, $%
(-\infty, x_1 - {\frac{\alpha |L| }{2}})$, $[x_1 - {\frac{\alpha |L| }{2}},
x_1 + {\frac{\alpha |L| }{2}}]$ and $(x_1 + {\frac{\alpha |L| }{2}}, \infty)$%
, where $\alpha$ is a positive constant to be fixed later. {\it The
absolute value $|L|$ used here allows negative $L$ values in the following
discussion.}

In the region
including the node one may perform the Taylor expansion $\phi(x)
\approx C_{1} (x-x_1)$ and so $p(x) \approx C_{1}^{2} (x-x_1)^2$. Thus 
\begin{eqnarray}
\lefteqn{ \delta E_{node} \approx } \nonumber \\
& & {\frac{\hbar^2 C_{1}^2 }{4 m L^2 \eta^4}} \int_{{\frac{%
-\alpha |L| }{2}}}^{{\frac{\alpha |L| }{2}} } dx x^2 \, \mbox{[} \ln {\frac{x^2 }{%
(1-\eta) x^2 + \eta (x + \eta L)^2 }} + 1  \nonumber \\
&& \;\;\;\;\;\;\;\;\;\;\;\;\;\;\;\;\;\;\;\;\;\;\;\;\;\;\;\;   - {\frac{(1-\eta) x^2 }{(1-\eta) x^2 + \eta (x + \eta L)^2 }} 
 - {\frac{\eta (x-\eta L)^2 }{(1-\eta) (x-\eta L)^2 + \eta x^2}} \mbox{]} \,  \label{EPO} \\
&& {} = {\frac{\hbar^2 C_{1}^2 |L| }{4 m \eta^4}} \int_{-\alpha /2}^{\alpha /2}
dy y^2 \, \mbox{[} \ln {\frac{y^2 }{(1-\eta) y^2 + \eta (y + \eta)^2 }} + 1 \nonumber \\
&& \;\;\;\;\;\;\;\;\;\;\;\;\;\;\;\;\;\;\;\;\;\;\;\;\;\;\;\;  - { \frac{(1-\eta) y^2 }{(1-\eta) y^2 + \eta (y + \eta )^2 }} - {\frac{\eta
(y-\eta )^2 }{(1-\eta) (y-\eta)^2 + \eta y^2}} \mbox{]} \, + O(L^2). \nonumber \\
&& \label{EP} 
\end{eqnarray}
Notice that the leading $|L|^{3}$ part
of the integral (\ref{EPO}) comes already from the $\int dxx^{2}$ piece after the
scaling $x=|L|y$, so subleading terms in the Taylor expansion of $%
p_{n}(x)\approx C_{np}^{2}(x-x_{p})^{2}+O(x^{3})$ contribute only at $%
O(L^{2})$ to $\delta E$.

For $\phi (x)$ having nodes at $x=x_{1},x_{2},....,x_{N}$ we may repeat the above procedure in the
neighbourhood of each node as long as the nodes are widely
separated. Then 
\begin{equation}
\delta E_{nodes} \approx {\frac{\hbar ^{2}|L|}{4m\eta ^{4}}}\ J(\eta ,\alpha )\
\sum_{p=1}^{N}C_{np}^{2}\,+ O(L^{2}) \label{EPP}
\end{equation}
with 

\begin{eqnarray}
\lefteqn{ J(\eta ,\alpha ) \equiv } \nonumber \\
& & \int_{{\frac{-\alpha }{2}}}^{{\frac{\alpha }{2}}} dy y^{2} \mbox{[} 
 \ln {\frac{y^{2}}{(1-\eta )y^{2}+\eta (y+\eta )^{2}}}+1 \nonumber \\
&& \;\;\;\;\;\;\;\;\;\;\;\;\;\;\;\;\;\;  -{\frac{(1-\eta )y^{2}}{(1-\eta )y^{2}+\eta (y+\eta )^{2}}}-{\frac{\eta (y-\eta )^{2}
}{(1-\eta )(y-\eta )^{2}+\eta y^{2}}} \mbox{]} \,.  \label{JJ} 
\end{eqnarray}
In (\ref{EPP}) $n$ refers to the quantum number(s) of the unperturbed
state and $p$ labels a node. 

To fix the value of $\alpha$ we have to look at the {\it nodeless} regions of the the first integral in (\ref{diff}): 
The terms $p(x \pm \eta L)$ may be safely expanded about $L=0$ to give for the {\it integrand} a series $\sim L^0+ L^2 +...$; there is no $L^1$ term in the series because of parity invariance (\ref{parity}). But since the integration limits are dependent on $\alpha |L|$ we have to be sure that the integral receives no enhancement of $1/|L|$ factors from them and therefore we need to choose 
\begin{equation}
\alpha = {C a \over |L|} \, , \label{alpha}
\end{equation}
where $C$ is a positive constant, so as to make the integration limits $L$-independent. Then, since $\delta E = 0$ for $L = 0$, the $O(L^0)$ piece from the nodeless regions of the first integral in (\ref{diff}) must cancel the second integral in (\ref{diff}) leaving a net contribution of order $L^2$. 

Returning now to (\ref{EPP}) and using (\ref{alpha}), we deduce that for small values of our perturbative parameter $|L|/a$ we need to expand $J(\eta, \alpha)$ for $\alpha$ large. 
We find
\begin{equation}
J(\eta, \alpha \to \infty) \to {\frac{-2 }{3}} \sqrt{1 - \eta} \
\eta^{9/2} \ (4 \eta -1) \pi + O \left( {1 \over \alpha} = {|L| \over C a} \right) \label{asymp} \, .
\end{equation}
Define the $\alpha$-independent piece of (\ref{asymp}) as 
\begin{equation}
J(\eta) \equiv {\frac{-2 }{3}} \sqrt{1 - \eta} \
\eta^{9/2} \ (4 \eta -1) \pi \, \label{J}
\end{equation}
so that one may finally write for (\ref{diff})
\begin{equation}
\delta E = {\frac{\hbar ^{2}|L|}{4m\eta ^{4}}} \ J(\eta) \
\sum_{p=1}^{N}C_{np}^{2}\, + O(L^{2}) \, . \label{EPF}
\end{equation} 

A remarkable aspect of the formula (\ref{EPF}) is that the 
specific dependence on the external potential $V(x)$ has been factorised:
it is only in the $C_{np}$ coefficients. Since $J(\eta)$ vanishes at $\eta = \eta_c =1/4$, it means that the leading energy shifts vanish at a universal, $V(x)$ independent, critical point!  

Given the intricate steps in above derivation, it is useful to perform some checks. From our
numerical investigations we found that for the excited states of the infinite well,
$\delta E =0$ for a value of $\eta$ between $(0.24, 0.25)$, in close agreement with the above 
prediction. We also confirmed numerically that the energy shifts for the excited
states of the SHO also vanish at essentially (limited by our numerical
precision) the same $\eta_c, 0.24-0.25$ as that for the infinite well. 
We remark that since the infinite well may be thought of as the $\gamma \to \infty$ limit of the potential $V(x) = |x/a|^{\gamma}$, so we are essentially checking (\ref{EPF}) at two limiting ends of a class of potentials.

More complete checks of (\ref{EPF}) involve comparing it with the $n$-dependent expressions for the energy shifts found numerically in the previous section. These checks are done in the Appendices, again showing good agreement.

Let us summarise some of the main features of (\ref{EPF}): Firstly, the expression clearly shows the non-analytic $O(|L|)$ trend confirmed numerically in the previous section. It also shows that the energy shift will be negative (positive) for large (small) $\eta $ values.

\subsection{Higher Dimensions}

For higher dimensions an explicit analysis similar to the preceeding subsection is
awkward because the nonlinearity is expressed in the preferred Cartesian
basis with broken rotational symmetry whereas most potentials, such as the
hydrogen atom, have a symmetry and so are better expressed in other
coordinate systems. Nevertheless, we can make some general statements.

For nodeless states we have the analog of (\ref{formal}) by expanding (\ref{3dd}) and
so get $\delta E\sim O(L^{2})$. 

For excited states the presence of nodes leads to singularities as before in
the naive Taylor expansion. Arguments similar to above then imply that $%
\delta E$ will be enhanced to $O(|L|)$ as each coordinate is
treated separately in (\ref{3dd}). Thus we expect again the energy shifts to be
positive for small $\eta$ and negative for larger $\eta$, vanishing at some
intermediate value. The numerical results of Sect.(3.3) for the angular
momentum states $l \neq n-1$ are in agreement with these general
expectations although we have not checked the expected variation with $\eta$.

A very interesting situation arises for the $l= n-1$ states of the hydrogen
atom for which the radial wavefunction vanishes only at the origin, 
\begin{equation}
\psi_{n,l=n-1} \sim r^{n-1} Y_{lm} (\theta, \phi) \,.
\end{equation}
Although the corresponding probability density $p(\vec{r})$ has a node at
the origin, the radial integral in the $O(L^2)$ contribution 
$\delta E \sim
\int d \Omega \int_{0}^{\infty} dr p(r) \left( ...\right)$, the three dimensional analog of (\ref{formal}),   is nonsingular,
as we see by power counting, if $2(n-1) + 3 > 4$, that is, for $n \ge 2$.
This explains the ``anomalous", $\delta E \sim L^2$, behaviour of such excited states observed in
Sect.(3.3).

\section{Exactly solved models}

In using perturbation theory we have assumed, as is usually done in physics, that the quantity
of interest will deform continuously as the perturbation is turned on. Here we
briefly discuss two {\it nonlinear} Schrodinger equations for which exact
solutions are available so that one can test perturbation theory. In
addition, the models will be used to further highlight some of the
distinctive features we have observed for the nonlinear equation (\ref{nsch1}). For simplicity we consider 
only the one dimensional case here.

\subsection{Gross-Pitaevskii (GP) Equation}

This classic \cite{sulem} equation is used as an effective theory in studies
of condensed matter. It corresponds to using $F(p(x,t))=gp(x,t)$ in (\ref
{nsch1}). To leading order, one has 
\begin{equation}
\delta E=g\int p^{2}dx\,  \label{pert-gp}
\end{equation}
so that energy shifts are always positive or negative depending on the sign
of the coupling $g$. Explicitly for the infinite well one has 
\begin{equation}
\delta E_n =\frac{3g}{2a},
\end{equation}
a constant shift independent of $n$, as obtained earlier in Ref.\cite
{gp-exact} which also showed that this perturbative result was the appropriate
limit of the exact solution of this equation with the infinite well
potential.

For the SHO potential we are not aware of any exact solutions for the GP
equation but (\ref{pert-gp}) gives the leading order correction, 
\begin{equation}
\delta E\approx \frac{g}{a\sqrt{2\pi }}n^{-0.31}, \label{sho-gp}
\end{equation}
for $n \ge 1$, showing that it decreases with $n$. We obtained (\ref{sho-gp}) through a numerical best-fit to an assumed power law.

The constant or decreasing dependence of the energy shifts on $n$,
respectively for the above two potentials in the GP equation, should be
contrasted with the results for the information-theoretic nonlinearity (\ref{FF}) which
showed an {\it increasing} dependence on $n$. As we saw in Sect.(4) that increasing dependence on $n$
was due to the prominent role played by nodes which by contrast are
completely irrelevant in (\ref{pert-gp}).

\subsection{A Pseudo-nonlinear model}

Starting from the usual linear Schrodinger equation 
\begin{equation}
i \hbar {\frac{\partial \Psi }{\partial t}} = - {\frac{{\hbar}^2 }{2m}} {%
\frac{\partial^2 \Psi }{\partial x^2}} + V(x) \Psi \, ,
\end{equation}
we can re-arrange the kinetic term by an amount $\epsilon$ to obtain 
\begin{equation}
i \hbar {\frac{\partial \Psi }{\partial t}} = - (1-\epsilon) {\frac{{\hbar}%
^2 }{2m}} {\frac{\partial^2 \Psi }{\partial x^2}} + V(x) \Psi - {\frac{
\epsilon }{\Psi}} {\frac{{\hbar}^2 }{2m}} \left( {\frac{\partial^2 \Psi }{%
\partial x^2}} \right) \Psi \, ,  \label{pseudo}
\end{equation}
which corresponds to an equivalent nonlinear Schrodinger equation with mass $%
m/(1-\epsilon)$ in the linear part and a perturbed nonlinearity 
\begin{equation}
F(p) \equiv - {\frac{ \epsilon }{\Psi}} {\frac{{\hbar}^2 }{2m}} \left( {%
\frac{\partial^2 \Psi }{\partial x^2}} \right) \, . \label{2nd}
\end{equation}
Thus in this case the exact and unperturbed solutions just correspond to a
mass renormalisation. For stationary states first order perturbation theory gives 
\begin{eqnarray}
\delta E &=& {\frac{-\epsilon {\hbar}^2 }{2m}} \int dx {\frac{p }{\sqrt{p}}} 
{\frac{\partial^2 \sqrt{p} }{\partial x^2}} \\
&=& {\frac{\epsilon {\hbar}^2 }{2m}} \int dx \left( {\frac{\partial \sqrt{p} 
}{\partial x}} \right)^2 \,  \label{pert-pseudo}
\end{eqnarray}
so that again the energy shifts are simply correlated in sign with the sign
of $\epsilon$.

As the wavefunctions for the linear Schrodinger equation with an infinite
well potential are independent of the mass and so also of $\epsilon$, the
first order correction using (\ref{pert-pseudo}) gives an exactly $\epsilon$
contribution in this case. When that is added to the unperturbed energies which are
proportional to $1-\epsilon$, one gets the full answer, that is, first order
perturbation theory for this problem is all there is. 

For the SHO and other
problems the first order correction will generally, by construction, lead
to final results correct up to errors of $O(\epsilon)^2$.

Thus in this nonlinear model the perturbative corrections to the energy always have the same $n$ dependence as that for the unperturbed energies. Therefore one may interpret the analogous results for Eq.(\ref{FF}) as due in some rough sense to higher derivative terms, higher than the second-order kinetic energy terms like (\ref{2nd}). This is indeed what is implied by a formal expansion of (\ref{FF}) but as we discussed in Sections (2,4), that formal expansion is in general singular and the actual result depends acutely on whether the unperturbed states do or do not have nodes.

\section{Conclusion}

Our main result is 
\begin{equation}
\delta E = {\frac{\hbar ^{2}|L| \pi}{6m}}\ \sqrt{\eta(1 - \eta)}\ \
(1-4 \eta)
\sum_{p=1}^{N}C_{np}^{2}\, + O(L/a)^2 \, , \label{EPX1}
\end{equation}
which gives the leading correction, due to the nonlinearity (\ref{FF}), to the energy eigenvalues of the usual $1+1$ dimensional linear Schrodinger equation for cases where the unperturbed states have nodes. The correction is propotional to $|L|/a$, hence it is non-anayltic and an enhancement over the correction for states without nodes for which $\delta E \propto L^2.$ The dependence of $\delta E$ on the external potential is only through the $C_{np}$ coefficients defined in Sect.(4.2). 

From (\ref{EPX1}) we see that independent of the external potential, $V(x)$, the leading energy correction to states with nodes vanishes at $\eta =1/4$. The existence of such a {\it universal} critical point is quite unexpected as neither the equations of motion nor  the full expression (\ref{delta}) indicate  such a special point.

As eq.(\ref{EPX1}) shows, $\delta E <0$ for $\eta > 1/4$, being positive for smaller $\eta$. Since $\eta$ is a free parameter in the nonlinear equation (\ref{nsch1}), it means that there is a qualitative difference in the properties of that equation for $\eta$ small or large. It is also interesting to note from (\ref{EPX1}) that the expression  is real precisely in the range $0< \eta <1$, which is exactly the explicit condition on $\eta$ we started with. Since for $\eta \to 0$ one formally has the linear theory, the square-root factor again emphasizes, in addition to the $|L|$ term, the non-analytic character of (\ref{EPX1}). 

For the usual linear Schrodinger equation in one space dimension, states with nodes are the excited states of a system although in some cases, such as that for the infinite well, the ground state also has nodes. For states without nodes, which are typically ground states, such as for the SHO, the leading energy corrections are of order $L^2$ and given by a simple expansion of (\ref{delta}). Thus at $\eta =1/4$ all the states of 
system, with or without nodes, have $\delta E \propto L^2$.

For higher dimensions, the qualitative properties are similar to the one dimensional case. That is, nodeless states 
get $\delta E \propto L^2$ while states with nodes in general have $\delta E \propto L$. We saw an exception in the hydrogen atom example where some excited states with nodes had $\delta E \propto L^2$ because the potential singularities were integrable. 

Let us now discuss the validity of perturbation theory for the infinite well and SHO where $\delta E$ increases with $n$, the principal quantum number, faster than the unperturbed states. For example, in the infinte well case we found $\delta E \propto |L| n^3 /a$ so that even if $|L|/a <<1$, at large $n$ the correction $\delta E$ would no longer be small. This indicates a breakdown of first order perturbation theory for large $n$ states, requiring one to go to higher orders. Presumably, if $L/a$ is small, the net perturbative correction should be small for all $n$, so one expects the higher order corrections to sum to a reasonable expression. A simple Pade resummation suggests $\delta E \propto n^2/(1 + b n)$ for the infinite well at $\eta =1/2$, where $b>0$ is some constant. 

Finally, we discuss some physical implications of our result if the nonlinearity (\ref{FF}) is a fundamental or effective representation of potential new physics at short distances as suggested in \cite{RP2}.  
For a particle in a large box, we may use the infinite well result, generalised {\it via} (\ref{3d}) to three independent dimensions, to see that for $\eta > 1/4$ high energy states have their energies lowered, that is the nonlinearity acts to moderate high energy divergences. One reaches the same conclusion from the SHO results if one thinks of ordinary free quantum field theory modes as SHO states. 

Thus the nonlinearity (\ref{FF}), applied here heuristically to field theory, suggests that the usual high energy divergences of quantum field theory might be moderated, if not absolutely eliminated. Now in \cite{RP1,RP2} it was suggested that the nonlinearity (\ref{FF}) might be linked to gravity simply by requiring $L$ to be a universal length scale. Taken together, this then suggests that gravity might moderate ultraviolet divergences of field theory. Interestingly, the suggestion that gravity might regulate ultraviolet divergences  has been made several times in the past through different reasoning within the context of usual linear quantum theory, see for example \cite{gravity} and references therein.

However the above moderation works only for $\eta > 1/4$ where we have $\delta E <0$ for excited states. What if in reality one has $\eta < 1/4$? Then $\delta E >0$ and this means that we are quite possibly under-estimating the amount of energy in quantum systems. One wonders if this might be relevant for the dark energy/matter problem in cosmology.  

So it appears that knowing the physically relevant value of $\eta$ is quite important for potential phenomenological applications of the nonlinear equation. Perhaps $\eta$ could be fixed theoretically through a renormalisation group study of a discretised version of the nonlinear equation (\ref{FF}). In this regard, the naturally induced discretisation noted in \cite{tan} might be useful.

\newpage

\section{Appendix A: Infinite Well Revisited}

For the infinite well we may evaluate $\delta E$ (\ref{EPF}) explicitly since the $%
C_{np} $ for these case are easily obtained from the wavefunctions 
(\ref{well}), 
\begin{equation}
C_{np}=\sqrt{{\frac{2}{a}}}\ {\frac{n\pi }{a}}\ (-1)^{p} \, , \,\,\,  0<p\leq n\,.
\label{cnp}
\end{equation}
 Thus the dimensionless
energy shift is 
\begin{eqnarray}
\delta \tilde{E} &=&{\frac{a^{2}|L|}{2\pi ^{2}\eta ^{4}}}\sum_{p=1}^{n}\left[
{\frac{2}{a}}\left( \frac{n\pi }{a}\right) ^{2}\right] J(\eta)
\label{well-del} \\
&=&{\frac{|L|}{a}}n^{3}{\frac{J(\eta)}{\eta ^{4}}}\,.
\end{eqnarray}
The formula {\it is} valid also for the ground state, $n=1$, because the corresponding wavefunction vanishes at the two end points, each of
which contributes the equivalent of half of one regular node as can be seen
by reviewing the derivation of (\ref{EPP}) above.
We therefore now have an understanding of the intriguing $n^{3}$ behaviour seen
numerically in Sect.(3): each $C_{np}^{2}$ contributes an identical $n^{2}$
piece to the sum over $n$ terms.

We find at $\eta =1/2$, 
\begin{eqnarray}
\delta \tilde{E}_n &=& {\frac{|L| }{a}} 16 n^3 J(1/2) \\
& =& - 1.05{\frac{|L| }{a}} n^3 \, ,
\end{eqnarray}
 in good agreement with our numerical study of the infinite well in Sect.(3) which indicated an average value of $1.03$ for the numerical factor.

For other potentials an explicit evaluation of the sum in ({\ref{EPP}) does
not appear feasible as the coefficients in general are very complicated
functions of $n$ and $p$ that are rarely known in closed form. However an
asymptotic or numerical evaluation of $\sum C_{np}^{2}$ might be possible if
an explicit dependence on $n$ is required. We illustrate this for the SHO in the next Appendix.

\section{Appendix B: Semi-analytical Analysis of SHO Energy Shifts}

Recall that our analytical estimates of the energy shifts for the excited
states gave 
\[
\delta E = {\frac{\hbar ^{2}|L|}{4m\eta ^{4}}}\ J(\eta ) \
\sum_{p=1}^{N}C_{np}^{2}\,+O(L^{2}).
\]
For the SHO, we obtain $C_{np}$ from the wavefunctions 
\[
\psi _{n}(z)=N_{n}(\pi a^{2})^{-1/4}H_{n}(z)\exp (-z^{2}/2),
\]
where $z=x/a$ , $N_{n}=1/\sqrt{2^{n}n!},$ and $H_{n}(z)$are the $n$-th order
Hermite polynomials. Observe that the wavefunction vanishes only when the
Hermite polynomial is zero.

What is required is the Taylor expansion of the wavefunction about the
nodes. At the nodes, 
\begin{equation}
H_{n}(z_{p}^{n})=0  \;\;\; \mbox{for} \;\;\;  p=1,2,...n,
\end{equation}
where $z_{p}^{n}$ refers to the $p$-th root of $H_{n}(z)$. Therefore near a root we have, to
leading order, 
\begin{eqnarray*}
\psi _{n}(z) &\approx &\psi _{n}(z_{p}^{n})+\frac{d\psi _{n}}{dz}%
(z_{p}^{n})(z-z_{p}^{n}) \\
&=&\frac{d\psi _{n}}{dz}(z_{p}^{n})(z-z_{p}^{n}).
\end{eqnarray*}
Reverting to $x$, one obtains 
\begin{equation}
C_{np}=\frac{d\psi _{n}}{dx}(z_{p}^{n})=\frac{1}{a}\left( N_{n}(\pi
a^{2})^{-1/4}H_{n}^{\prime }(z_{p}^{n})\exp \left[ -\left( z_{p}^{n}\right)
^{2}/2\right] \right) .
\end{equation}
Using the identity 
\[
H_{n+1}(x)+H_{n}^{\prime }(x)=2xH_{n}(x)
\]
then gives 
\begin{equation}
H_{n}^{\prime }(z_{p}^{n})=-H_{n+1}(z_{p}^{n}).
\end{equation}
Finally, 
\begin{equation}
C_{np}^{2}=\frac{1}{a^{3}}\frac{1}{\sqrt{\pi }}\frac{1}{2^{n}n!}\left[
H_{n+1}(z_{p}^{n})\right] ^{2}\exp \left( -(z_{p}^{n})^{2}\right) .
\end{equation}
We evaluated the sum of these $C_{np}^{2}$ through a numerical computation of
the roots and sums of the Hermite polynomials. Since the `$a$'
behaviour is already explicit, we examined the $n$-dependence by calculating
$\sum $ $a^{3}C_{np}^{2}$ from $n=1$ to $n=23$. Furthermore at  
$\eta =1/2$: 
\begin{eqnarray}
\delta \tilde{E} &=&\frac{a^{2}}{4}{\frac{|L|}{\eta ^{4}}}\left( \sum
C_{np}^{2}\ \right) J(\eta=1/2) \\
&=& -0.27 {\frac{|L|}{a}} n^{1.40} \, .
\end{eqnarray}
This result is in good agreement with the purely numerical one quoted in Section (3) (which was actually for relatively low values of $a$).

\newpage

\section*{Figure Captions}

\begin{itemize}
\item  Figure 1 : Log-log plot of $| \delta \tilde{E}|$ vs. $%
a$ for the ground state $n=1$ of the infinite well. The line has slope $%
0.99985(2)$ and intercept $0.0448(1)$. In this and the other figures the `logs' are natural logarithms.

\item  Figure 2 : Log-log plot of $\left| \delta \tilde{E}\right| $ vs. $n$
for $a=1000$ of the SHO. The line has slope $1.413(8)$ and intercept $%
-8.25(2)$.

\item  Figure 3 : Plot of $|\delta E/E_{0}|$ vs. $n\geq 2$ for $l=0$ states
of Hydrogen atom. Curves for different values for $a$ are shown.

\end{itemize}

\vspace{2cm}

\section*{Figures}
\begin{figure}[h]
  \begin{center}
   \epsfig{file=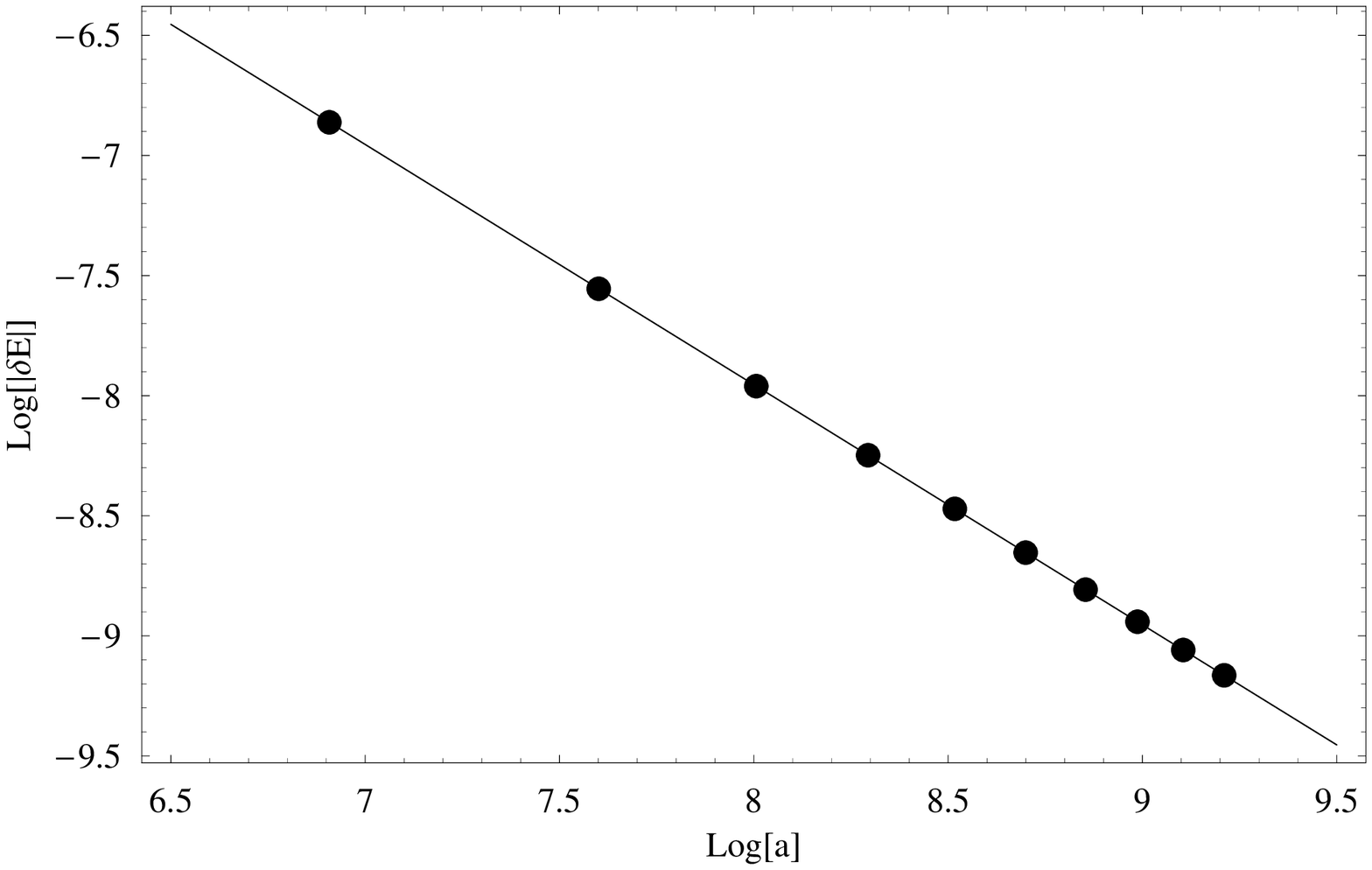, width=9cm}
    \caption{.}
    \label{energy}
  \end{center}
\end{figure}

\begin{figure}[h]
  \begin{center}
   \epsfig{file=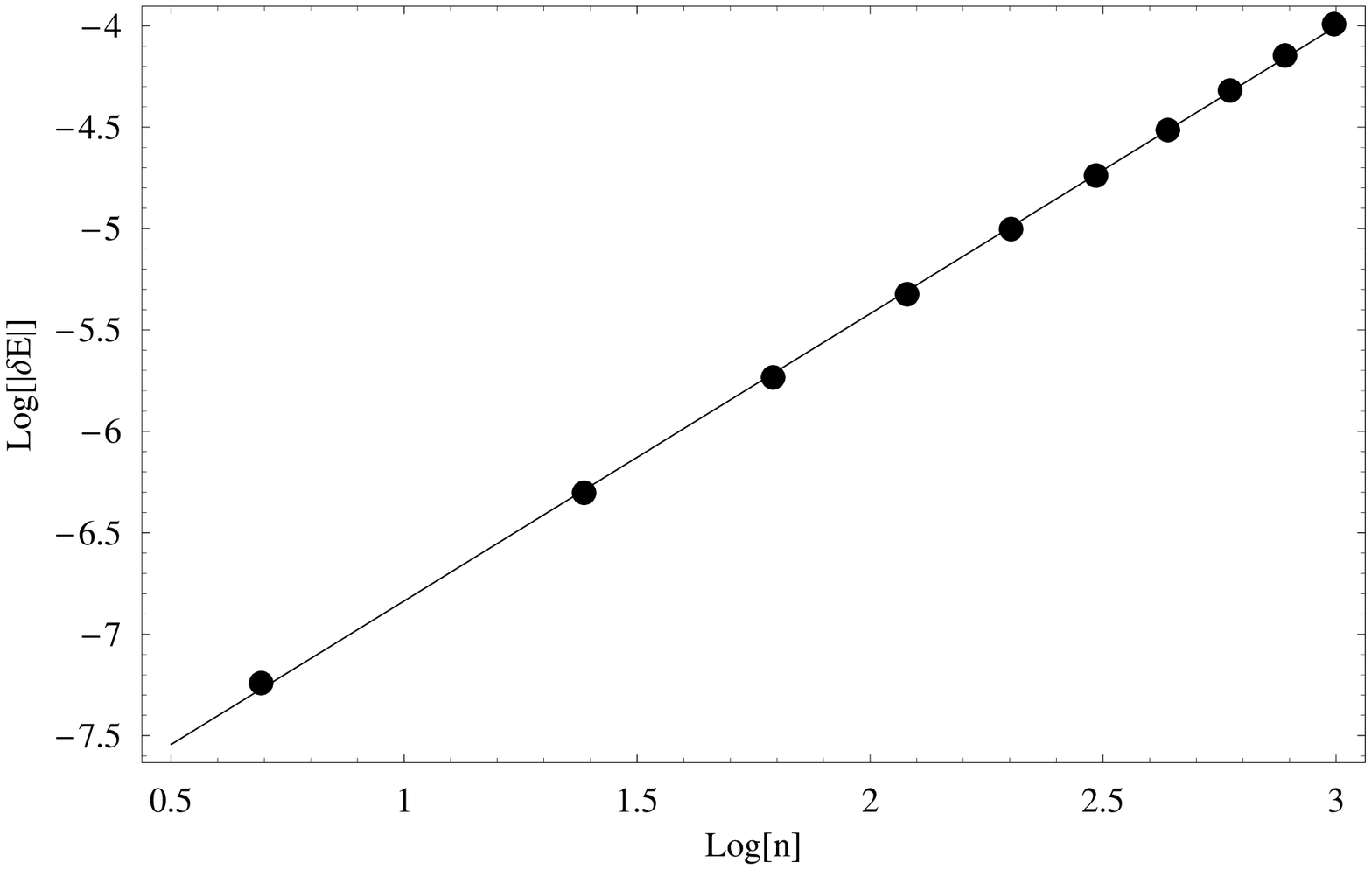, width=9cm}
    \caption{.}
    \label{energy}
  \end{center}
\end{figure}

\begin{figure}
  \begin{center}
   \epsfig{file=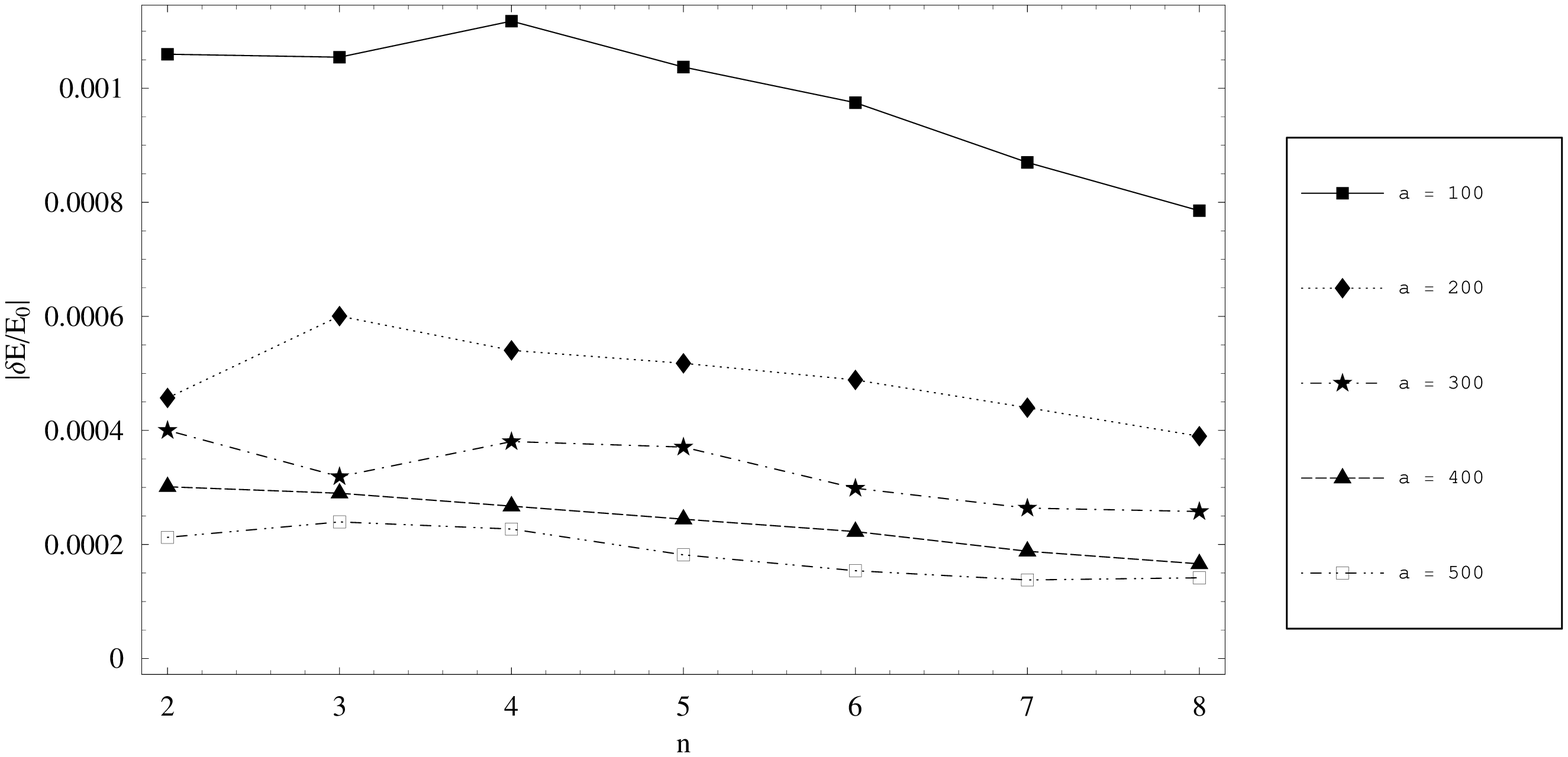, width=9cm}
    \caption{.}
    \label{energy}
  \end{center}
\end{figure}

\end{document}